\documentclass{llncs}
\usepackage{makeidx}  
\usepackage{graphicx}        
\usepackage{fancyvrb}
\usepackage{fixltx2e}
\usepackage{amsfonts}
\usepackage{alltt}
\usepackage[polish, german, english]{babel}
\begin{document}
\title{Communication-aware algorithms for target tracking in wireless sensor networks}
\titlerunning{Communication-aware algorithms for target tracking in wireless sensor networks}  
%
\selectlanguage{polish}
\author{Bart"lomiej P"laczek}
\authorrunning{Bart"lomiej P"laczek} 
\institute{Institute of Computer Science, University of Silesia,\\
B"edzi"nska 39, 41-200 Sosnowiec, Poland\\
\email{placzek.bartlomiej@gmail.com}
}

\maketitle              

\selectlanguage{english}

\begin{abstract}
This paper introduces algorithms for target tracking in wireless sensor networks (WSNs) that enable reduction of data communication cost. The objective of the considered problem is to control movement of a mobile sink which has to reach a moving target in the shortest possible time. Consumption of the WSN energy resources is reduced by transferring only necessary data readings (target positions) to the mobile sink. Simulations were performed to evaluate the proposed algorithms against existing methods. The experimental results confirm that the introduced tracking algorithms allow the data communication cost to be considerably reduced without significant increase in the amount of time that the sink needs to catch the target. \selectlanguage{polish}\footnote{Preprint of: P"laczek B.: Communication-Aware Algorithms for Target Tracking in Wireless Sensor Networks. Communications in Computer and Information Science, vol. 431, pp. 69-78 (2014). The final publication is available at www.springerlink.com}
\end{abstract}
%
\section{Introduction}
Wireless sensor networks (WSNs) can be utilized as target tracking systems that detect a moving target, localize it and report its location to the sink. So far, the WSN-based tracking systems have found various applications, such as battlefield monitoring, wildlife monitoring, intruder detection, and traffic control \cite{bplaczek:bib1,bplaczek:bib2}.

This paper deals with the problem of target tracking by a mobile sink which uses information collected from sensor nodes to catch the target. Main objective of the considered system is to minimize time to catch, i.e., the number of time steps in which the sink reaches the moving target. Moreover, due to the limited energy resources of WSN, also the minimization of data communication cost (hop count) is taken into consideration. It is assumed in this study that the communication between sensor nodes and the sink involves multi-hop data transfers.

Most of the state-of-the-art data collection methods assume that the current location of the target has to be reported to sink continuously with a predetermined precision. These continuous data collection approaches are not suitable for developing the WSN-based target tracking applications because the periodical transmissions of target location to the sink would consume energy of the sensor nodes in a short time. Therefore, the target tracking task requires dedicated algorithms to ensure the amount of data transmitted in WSN is as low as possible.

Intuitively, there is a trade-off between the time to catch minimization and the minimization of data communication cost. In this study two algorithms are proposed that enable substantial reduction of the data collection cost without significant increase in time to catch. The introduced communication-aware algorithms optimize utilization of the sensor node energy by selecting necessary data readings (target locations) that have to be transmitted to the mobile sink. Simulation experiments were conducted to evaluate the proposed algorithms against state-of-the-art methods. The experimental results show that the presented algorithms outperform the existing solutions.

The paper is organized as follows. Related works are discussed in Section 2. Section 3 contains a detailed description of the proposed target tracking methods. The experimental setting, compared algorithms and simulation results are presented in Section 4. Finally, conclusion is given in Section 5.
%
\section{Related works}
In the literature, there is a variety of approaches available that address the problem of target tracking in WSNs. However, only few publications report the use of WSN for chasing the target by a mobile sink. Most of the previous works have focused on delivering the real-time information about trajectory of a tracked target to a stationary sink. This section gives references to the WSN-based tracking methods reported in the literature that deal explicitly with the problem of target chasing by a mobile sink. A thorough survey of the literature on WSN-based object tracking methods can be found in references \cite{bplaczek:bib1,bplaczek:bib4}.

Kosut et al. \cite{bplaczek:bib5} have formulated the target chasing problem, which assumes that the target performs a simple random walk in a two-dimensional lattice, moving to one of the four neighbouring lattice points with equal probability at each time step. The target chasing method presented in \cite{bplaczek:bib5} was intended for a system composed of static sensors that can detect the target, with no data transmission between them. Each static sensor is able to deliver the information about the time of the last target detection to the mobile sink only when the sink arrives at the lattice point where the sensor is located.

A more complex model of the WSN-based target tracking system was introduced by Tsai et al. \cite{bplaczek:bib6}. This model was used to develop the dynamical object tracking protocol (DOT) which allows the WSN to detect the target and collect the information on target track. The target position data are transferred from sensor nodes to a beacon node, which guides the mobile sink towards the target. A similar method was proposed in \cite{bplaczek:bib7}, where the target tracking WSN with monitor and backup sensors additionally takes into account variable velocity and direction of the target.

In this paper two target tracking methods are proposed that contribute to performance improvement of the above-mentioned target tracking approaches by reducing both the time to catch (i.e., the time in which mobile sink can reach the target) and the data communication costs in WSN. In this study, the total hop count is analysed to evaluate the overall cost of communications, however it should be noted that different metrics can also be also used, e.g., number of data transfers to sink, number of queries, number of transmitted packets, and energy consumption in sensor nodes.

The introduced algorithms provide decision rules to optimize the amount of data transfers from sensor nodes to sink during target chasing. The research reported in this paper is a continuation of previous works on target tracking in WSN, where the data collection was optimized by using heuristic rules \cite{bplaczek:bib9} and the uncertainty-based approach \cite{bplaczek:bib10}. The algorithms proposed in that works have to be executed by the mobile sink. In the present study the data collection operations are managed by distributed sensor nodes.

To reduce the number of active sensor nodes the proposed algorithms adopt the prediction-based tracking method \cite{bplaczek:bib11}. According to this method a prediction model is applied, which forecasts the possible future positions of the target. On this basis only the sensor nodes expected to detect the target are activated at each time step.

\section{Proposed methods}
In this section two methods are proposed that enable reduction of data transfers in WSN during target tracking. The WSN-based target tracking procedure is executed in discrete time steps. At each time step both the target and the sink move in one of the four directions: north, west, south or east. Their maximum velocities (in segments per time step) are assumed to be known. Movement direction of the target is random. For sink the direction is decided on the basis of information delivered from WSN. During one time step the sink can reach the nearest segments $(x_S, y_S)$ that satisfy the maximum velocity constraint: $|x_S-x'_S| + |y_S-y'_S| \leq v_{max}$, where coordinates $(x'_S, y'_S)$ describe previous position of the sink. Sink moves into segment $(x_S, y_S)$ for which the Euclidean distance $d[(x_S, y_S), (x_D, y_D)]$ takes minimal value. Note that $(x_D, y_D)$ are the coordinates of target that were lately reported to the sink.

Let $(x_C, y_C)$ denote coordinates of the segment where the target is currently detected. The sensor node that detects the target will be referred to as the target node. According to the proposed methods the information about target position is transmitted from the target node to the sink only at selected time steps. If this information is transmitted then the destination coordinates at sink $(x_D, y_D)$ are updated, i.e, $(x_D, y_D) = (x_C, y_C)$. It means that the current position of the target is available for sink only at selected time steps. In remaining time periods the sink moves toward the last reported target position, which is determined by coordinates $(x_D, y_D)$.

Hereinafter, symbol $dir(x, y)$ will be used to denote the direction chosen by sink when moving toward segment $(x, y)$. At each time step, the coordinates $(x_D, y_D)$ and $(x_C, y_C)$ are known for the target node. Therefore, the target node can determine the direction which will be chosen by the sink in both cases: if the current target position is transmitted to the sink and if the data transfer is skipped.

According to the first proposed method, the coordinates $(x_C, y_C)$ are transmitted to the sink only if $dir(x_D, y_D) \neq dir(x_C, y_C)$, i.e., if the direction chosen on the basis of coordinates $(x_D, y_D)$ is different than the one selected by taking into account the current position $(x_C, y_C)$.

In the second proposed method, the target node evaluates probability $P[dir]$ that the move of sink in direction $dir$ will minimize its distance to the segment in which the target will be caught. The target coordinates $(x_C, y_C)$ are transferred to the sink only if the difference $P[dir(x_C, y_C)]-P[dir(x_D, y_D)]$ is above a predetermined threshold. To evaluate probabilities $P[dir]$, the target node determines an area where the target can be caught. This area is defined as a set of segments:
\begin{equation}
   A=\{(x,y):t_T(x,y) \leq t_S(x,y)\},
\end{equation}
where $t_T(x, y)$ and $t_S(x, y)$ are the minimum times required for target and sink to reach segment $(x, y)$.

Let $(x_S, y_S)_C$ and $(x_S, y_S)_D$ denote the segments into which the sink will enter at the next time step if it will move in directions $dir(x_C, y_C)$ and $dir(x_D, y_D)$ respectively. In area $A$ two subsets of segments are distinguished: subset $A_C$ that consists of segments that are closer to $(x_S, y_S)_C$ than to $(x_S, y_S)_D$ and subset $A_D$ of segments that are closer to $(x_S, y_S)_D$ than to $(x_S, y_S)_C$:
\begin{equation}
   A_C=\{(x,y):(x,y) \in A \wedge d[(x,y),(x_S,y_S)_C] < d[(x,y),(x_S,y_S)_D]\},
\end{equation}
\begin{equation}
   A_D=\{(x,y):(x,y) \in A \wedge d[(x,y),(x_S,y_S)_D] < d[(x,y),(x_S,y_S)_C]\},
\end{equation}
On this basis the probabilities $P[dir]$ are calculated as follows:
\begin{equation}
  P[dir(x_C,y_C)]= \frac{|A_C|}{|A|}, \; P[dir(x_D,y_D)]= \frac{|A_D|}{|A|},
\end{equation}
where $|\cdotp|$ denotes cardinality of the set.

\begin{figure}
\centering
\includegraphics [width=6cm] {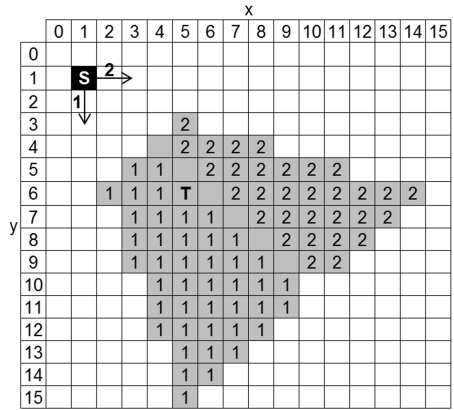}
\caption{Example of $P[dir]$ calculations}
\end{figure}
 
The operations discussed above are illustrated by the example in Fig. 1, where the positions of target and sink are indicated by symbols $T$ and $S$ respectively. Velocity of the target is 1 segment per time step. For sink the velocity equals 2 segments per time step. Gray color indicates the area $A$ in which the sink will be able to catch the target. The direction $dir(x_C, y_C)$ is shown by the arrow with number 1 and $dir(x_D, y_D)$ is indicated by the arrow with number 2, thus $(x_S, y_S)_C = (1, 3)$ and $(x_S, y_S)_D = (3, 1)$. Subset $A_C$ includes gray segments that are denoted by 1. The segments with label 2 belong to $A_D$. In the analyzed example $|A| = 82$, $|A_C| = 44$, and $|A_D| = 31$. According to Eq. (4) $P[dir(x_C, y_C)] = 0.54$, $P[dir(x_D, y_D)] = 0.38$ and the difference of these probabilities equals to 0.16. 

If the first proposed method is applied for the analysed example then the data transfer to sink will be executed, since $dir(x_D, y_D) \neq dir(x_C, y_C)$, as shown by the arrows in Fig, 1. In case of the second method, the target node will send the coordinates $(x_C, y_C)$ to the sink provided that the difference of probability (0.16) is higher than a predetermined threshold. The threshold value should be interpreted as a minimum required increase in the probability of selecting the optimal movement direction, which is expected to be obtained after transferring the target position data.

\section{Experiments}
Experiments were performed in a simulation environment to compare performance of the proposed methods against state-of-the-art approaches. The comparison was made by taking into account two criteria: time to catch and hop count. The time to catch is defined as the number of time steps in which the sink reaches the moving target. Hop count is used to evaluate the cost of data communication in WSN.
\subsection{Experimental setting}
In the experiments, it was assumed that the monitored area is a square of 200 x 200 segments. Each segment is equipped with a sensor node that detects presence of the target. Thus, the number of sensor nodes in the analysed WSN equals 40 000. Communication range of each node covers the eight nearest segments. Maximum velocity equals 1 segment per time step for the target, and 2 segments per time step for the sink.

Experiments were performed using simulation software that was developed for this study. The results presented in Sect. 4.3 were registered for 10 random tracks of the target (Fig. 2). Each simulation run starts with the same location of both the sink (5, 5) and the target (100, 100). During simulation the hop counts are calculated assuming that the shortest path is used for each data transfer to sink, the time to catch is measured in time steps of the control procedure. The simulation stops when target is caught by the sink. 
\subsection{Compared algorithms}
In the present study, the performance is analysed of four WSN-based target tracking algorithms. Algorithms 1 and 2 are based on the approaches that are available in literature, i.e. the prediction-based tracking and the dynamical object tracking. These algorithms were selected as representative for the state-of-the-art solutions in the WSN-based systems that control the movement of a mobile sink which has to reach a moving target. The new proposed methods are implemented in Algorithms 3 and 4. The pseudocode in Tab. 1 shows the operations that are common for all the examined algorithms. Each algorithm uses different condition to decide if current position of the target will be transmitted to the sink (line 6 in the pseudocode). These conditions are specified in Tab. 2.

\begin{figure}
\centering
\includegraphics [width=12cm] {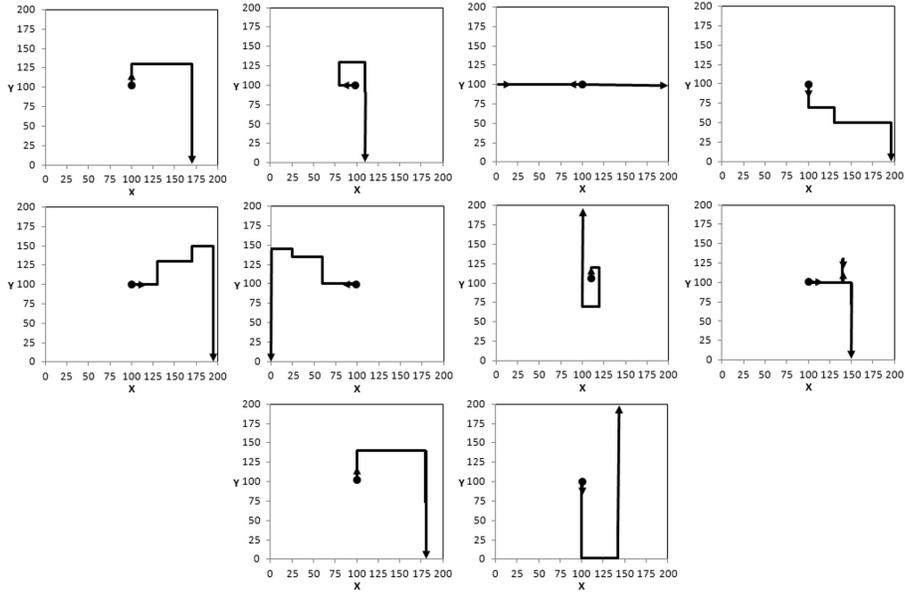}
\caption{Simulated tracks of target}
\end{figure}

For all considered algorithms, the prediction-based approach is used to select the sensor nodes that have to be activated at a given time step $(t)$. Prediction of the possible target locations is based on a simple movement model, which takes into account the assumptions on target movement directions and its maximum velocity. If for previous time step $(t-1)$ the target was detected in segment $(x'_C, y'_C)$, then at time step $t$ the set of possible target locations $M$ can be determined as follows: 
\begin{equation}
   M=\{(x,y): |x-x'_C|+|y-y'_C| \leq v_{max}\},
\end{equation}
where $v_{max}$ is the maximum velocity of target in segments per time step.

\begin{table}
\centering
\caption{Pseudocode for WSN-based target tracking algorithms}
\begin{tabular}{ll}
\hline
 1 & repeat \\ 
 2 & \hspace{0.3cm} at target node do \\ 
 3 & \hspace{0.6cm} determine $M$ \\ 
 4 & \hspace{0.6cm} collect data from each node$(x, y): (x, y) \in M$ \\ 
 5 & \hspace{0.6cm} determine $(x_C, y_C)$ \\ 
 6 & \hspace{0.6cm} if $condition$ then communicate $(x_C, y_C)$ to the sink \\ 
 7 & \hspace{0.6cm} set node$(x_C, y_C)$ to be target node \\ 
 8 & \hspace{0.3cm} at sink do  \\ 
 9 & \hspace{0.6cm} if data received from target node then $(x_D, y_D) := (x_C, y_C)$ \\ 
 10 & \hspace{0.6cm} move toward $(x_D, y_D)$ \\ 
 11 & until $(x_S, y_S)=(x_C, y_C)$ \\ 
\hline 
\end{tabular} 
\end{table}

Algorithm 1 uses the prediction-based tracking method without any additional data transfer condition. According to this algorithm, the target location is reported to the sink at each time step. Sensor nodes for all possible target locations $(x, y) \in M$ are activated, and the discovered target location is transmitted to the sink. An important feature of Algorithm 1 is that the information about current target position is delivered to the sink with the highest available frequency (at each time step of the tracking procedure).

\begin{table}
\centering
\caption{Compared algorithms}
\begin{tabular}{lll}
\hline
Alg. &	Method & Data transfer condition \\
\hline
1 & Prediction-based tracking & None \\
2 & Dynamical object tracking & $(x_S, y_S)=(x_D, y_D)$ \\
3 & Proposed method \#1 & $dir(x_D, y_D) \neq dir(x_C, y_C)$ \\
4 & Proposed method \#2 & $P[dir(x_C, y_C)]-P[dir(x_D, y_D)] > threshold$ \\
\hline 
\end{tabular} 
\end{table}

Algorithm 2 is based on the tracking method which was proposed for the dynamical object tracking protocol. According to this approach sink moves toward location of so-called beacon node $(x_D, y_D)$. A new beacon node is set if the sink enters segment $(x_D, y_D)$. In such case, the sensor node which currently detects the target in segment $(x_C, y_C)$, becomes new beacon node and its location is communicated to the sink. When using this approach, the cost of data communication in WSN can be reduced because the data transfers to sink are executed less frequently than for the prediction-based tracking method.
The proposed communication-aware tracking methods are applied in Algorithm 3 and Algorithm 4 (see Tab. 2). Details of these methods were discussed in Sect. 3.
\subsection{Simulation results}
Simulation experiments were carried out in order to determine time to catch values and hop counts for the compared algorithms. As it was mentioned in Sect. 3, the simulations were performed by taking into account ten different tracks of the target. Average results of these simulations are shown in Fig. 3. It is evident that the best results were obtained for Algorithm 4, since the objective is to minimise both the time to catch and the hop count. It should be noted that Fig. 3. presents the results of Algorithm 4 for different threshold values. The relevant threshold values between 0.0 and 0.9 are indicated in the chart by the decimal numbers. According to these results, the average time to catch increases when the threshold is above 0.2. For the threshold equal to or lower than 0.2 the time to catch takes a constant minimal value. The same minimal time to catch is obtained when using Algorithm 3, however in that case the hop count is higher than for Algorithm 4. 

In comparison with Algorithm 1 both proposed methods enables a considerable reduction of the data communication cost. The average hop count is reduced by 47\% for Algorithm 3 and by 87\% for Algorithm 4 with threshold 0.2. Algorithm 2 also reduces the hop count by about 87\% but it requires much longer time to catch the target. The average time to catch for Algorithm 2 is increased by 52\%.
\begin{figure}
\centering
\includegraphics [width=7cm] {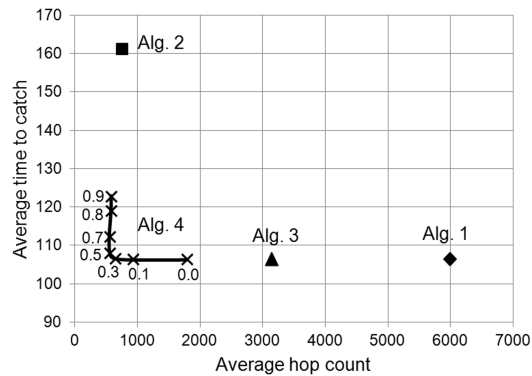}
\caption{Average time to catch and hop count for compared algorithms}
\end{figure}
\begin{figure}
\centering
\includegraphics [width=12cm] {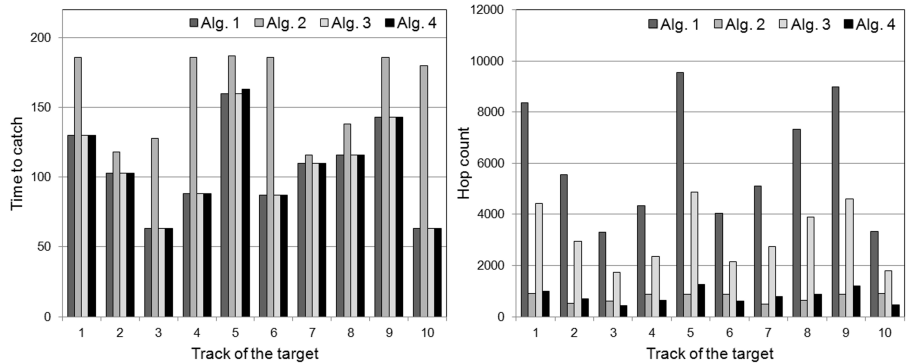}
\caption{Time to catch and hop counts for different tracks of target}
\end{figure}

Detailed simulation results are presented in Fig. 4. These results demonstrate the performance of the four examined algorithms when applied to ten different tracks of the target. The threshold value in Algorithm 4 was set to 0.2. The shortest time to catch was obtained by Algorithms 1, 3 and 4 for all tracks except the 5th one. In case of track 5, when using Algorithm 4 slightly longer time was needed to catch the target. For the remaining tracks the three above-mentioned algorithms have resulted in equal values of the time to catch. In comparison with Algorithm 1, the proposed algorithms (Algorithm 3 and Algorithm 4) significantly reduce the data communication cost (hop count) for all analysed cases. For each considered track Algorithm 2 needs significantly longer time to reach the moving target than the other algorithms. The hop counts for Algorithm 2 are close to those observed in case of Algorithm 4.
 
According to the presented results, it could be concluded that Algorithm 4, which is based on the proposed method, outperforms the compared algorithms. It enables a significant reduction of the data communication cost. This reduction is similar to that obtained for Algorithm 2. Moreover, the time to catch for Algorithm 4 is as short as in case of Algorithm 1, wherein the target position is communicated to the sink at each time step. 

\section{Conclusion}
The cost of data communication in WSNs has to be taken into account when designing algorithms for WSN-based systems due to the finite energy resources and the bandwidth-limited communication medium. In order to reduce the utilization of WSN resources, only necessary data shall be transmitted to the sink. This paper is devoted to the problem of transferring target coordinates from sensor nodes to a mobile sink which has to track and catch a moving target. The presented algorithms allow the sensor nodes to decide when data transfers to the sink are necessary for achieving the tracking objective. According to the proposed algorithms, only selected data are transmitted that can be potentially useful for reducing the time in which the target will be reached by the sink.

Performance of the proposed algorithms was compared against state-of-the-art approaches, i.e., the prediction-based tracking and the dynamical object tracking. The simulation results show that the introduced algorithms outperforms the existing solutions and enable substantial reduction in the data collection cost (hop count) without significant decrease in the tracking performance, which was measured as the time to catch.

The present study considers an idealistic WSN model, where the information about current position of target $(x_C, y_C)$ is always successfully delivered through multi-hop paths to the sink and the transmission time is negligible. In order to take into account uncertainty of the delivered information, the precise target coordinates $(x_C, y_C)$ should be replaced by a (fuzzy) set. Relevant modifications of the presented algorithms will be considered in future experiments. 

Although the proposed methods consider a simple case with a single sink and a single target, they can be also useful for the compound tracking tasks with multiple targets and multiple sinks \cite{bplaczek:bib3,bplaczek:bib8}. Such tasks need an additional higher-level procedure for coordination of the sinks, which has to be implemented at a designated control node, e.g., a base station or one of the sinks. The extension of the presented approach to tracking of multiple targets in complex environments is an interesting direction for future works.
%
%

\end{document}